\documentclass[
    ,final            % use final for the camera ready runs
  ]
  {aipproc}

\layoutstyle{8x11double}

\begin{document}

\title{Timing and Precession of the Young, Relativistic Binary Pulsar PSR J1906+0746}

%\classification{<Replace this text with PACS numbers; choose from this list:
 %               \texttt{http://www.aip.org/pacs/index.html}>}

\author{Laura Kasian}{
  address={University of British Columbia}
}

\author{PALFA Consortium}{
  address={see http://www.naic.edu/~palfa for complete list of participants}
%  ,altaddress={<author1 address>} % additional visiting address
}

\begin{abstract}

We present an updated timing solution and an analysis of the profile evolution - including precession and beam shape - of the young, relativistic binary pulsar J1906+0746. The 144-ms pulsar, in a 3.98-hour orbit with eccentricity 0.085 \cite{lorimer}, was initially discovered during the early stages of the ALFA (Arecibo L-band Feed Array) pulsar survey \cite{cordes} using the 305-metre Arecibo telescope and was subsequently found in archival Parkes Multibeam Survey data. We have since been regularly monitoring the system using the Arecibo and Green Bank telescopes, and include data from the Jodrell Bank, Parkes, Nancay and Westerbork telescopes. The nature of the binary companion will also be discussed based on improved estimates of the total and companion masses obtained from the updated timing solution. 

\end{abstract}

%%%%%%%%%%%%%%%%%%%%%%%%%%%%%%%%%%%%%%%%%%%%%%%%%%%%%%%%%%%%%%%%%%%
%%
%% The below \maketitle command inserts the actual front matter data.
%% It has to follow the above declarations.
%%
%%%%%%%%%%%%%%%%%%%%%%%%%%%

\maketitle

%%%%%%%%%%%%%%%%%%%%%%%%%%%%%%%%%%%%%%%%%%%%
%% MAINMATTER
%%
%%%%%%%%%%%%%%%%%%%%%%%%%%%%%%%%%%%%%%%%%%%%%%%%%%%%%%%%%%%%%%%%%%%%%%%%%%%%
%% Headings:
%%
%% The aipproc supports three heading levels, i.e., \section,
%%	\subsection, and \subsubsection.
%%%%%%%%%%%%%%%%%%%%%%%%%%%%%%%%%%%%%%%%%%%%%%%%%%%%%%%%%%%%%%%%%%%%%%%%%%%%
%% Cross-references:
%%
%% Page numbers (\pageref) and headings can NOT be referenced in the class,
%% since before being produced, no page numbers are determined.
%%
%% Tables, figures, and equeations can be referenced by using the LaTex
%% 	commands \label and \ref. For references to equation numbers, \eqref
%%	can be used, which will print "(1)" (while \ref will result in "1").
%%
%%%%%%%%%%%%%%%%%%%%%%%%%%%%%%%%%%%%%%%%%%%%%%%%%%%%%%%%%%%%%%%%%%%%%%%%%%%%
%% Lists: 
%%
%% Standard "itemize", "enumerate", etc. list environments are supported.
%%%%%%%%%%%%%%%%%%%%%%%%%%%%%%%%%%%%%%%%%%%%%%%%%%%%%%%%%%%%%%%%%%%%%%%%%%%%
%% Urls:
%%
%% \url{} command is provided for documenting URLs.
%%%%%%%%%%%%%%%%%%%%%%%%%%%%%%%%%%%%%%%%%%%%

\section{Gaussian Fitting and Profile Evolution}

In order to better understand the profile evolution and in turn remove some timing noise from the system, we have created standard profiles for each epoch of data, and fitted sets of 1,2, or 3 (as necessary) gaussians to each of the pulse and interpulse (seperately). All fitting was done using the gaussian fitting package bfit \cite{kramer}. After experimenting with several different methods of aligning the gaussian templates, we decided that the smoothest alignment was achieved by keeping the phase of the tallest gaussian constant. This also seems to produce a fairly monotonic behaviour in the phase of the interpulse. The full collection of modelled profiles is shown in Figure 1. These were used as standard profiles for timing.

\begin{figure}
  \includegraphics[height=.3\textheight]{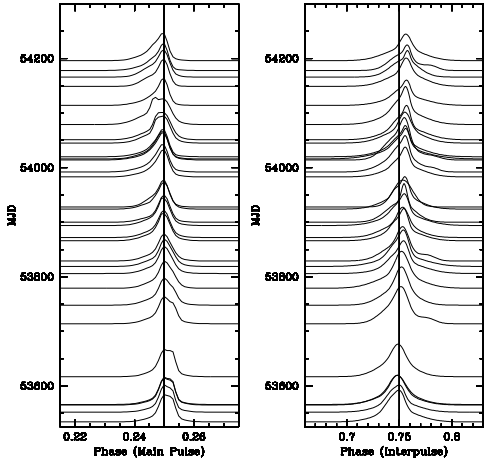}
  \caption{The fits of the pulse and interpulse of J1906+0746, based on ASP (Arecibo; coherent dedispersion), GASP (GBT; coherent dedispersion) and WAPP (Arecibo; spectrometer) observations, each modelled with several gaussian components. The pulses are spaced vertically based on observation date, and the interpulse is magnified by a factor of ten relative to the main pulse. Different machine dispersion methods account for some but not all of the apparent interpulse differences in nearby observations; the re maining differences are under investigation. The vertical line for the main pulse (left) illustrates the chosen alignment, whereas the vertical line for the interpulse demonstrates the marked phase shift of the interpulse relative to the main pulse (see proceedings by Desvignes et al.)}

\end{figure}

\section{Template Evolution}
The changing profile shape poses a problem in the creation of times-of-arrival (TOAs) for the system, as the changing profile shape allows for ambiguity in the fiducial point of the profile, which translates into additional timing noise for the system.  Figure 2 shows the changing fiducial point of the gaussian profiles over time.  In order to reduce this source of timing noise, we have decided to 1) use a series of standard profiles developed from the well-modelled epochs of ASP, GASP and WAPP data, and 2) align the Gaussian templates as described in the previous section.

\begin{figure}
  \includegraphics[height=.19\textheight]{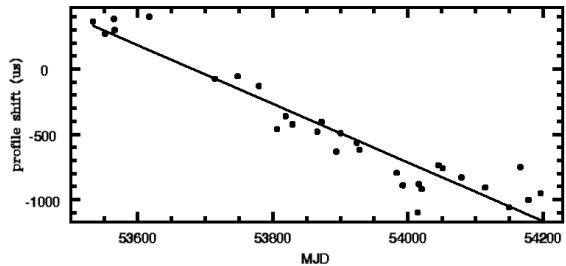}
  \caption{Phase offset of the profile at each of 29 epochs of data. Offsets computed by comparing one epoch of data with each of 29 gaussian-modelled standard profiles (as described in 'Gaussian Fitting and Profile Evolution' section).  A regression line is plotted through the points, and although there is a fair amount of scatter in the data, it roughly follows a linear trend.}
\end{figure}

\section{Orbital Aberration}
We obtained GBT time over four days between October 4 and 12, 2006, in an intensive campaign that has allowed us to study the effect of orbital aberration without the effects of the long-term profile evolution. Over such a small time span, the secular evolution of the pulse profile is negligible, and so we can be confident that any changes over orbital phase bins (as shown in Figure 3) are due solely to the change in observed emission cone direction as the pulsar moves at a relativistic orbital velocity.

\begin{figure}[h]
%  \begin{tabular}{cc}
%    \includegraphics[height=.3\textheight]{second} &
%    \includegraphics[height=.3\textheight]{first} 
  \begin{tabular}{c}
    \includegraphics[height=.3\textheight]{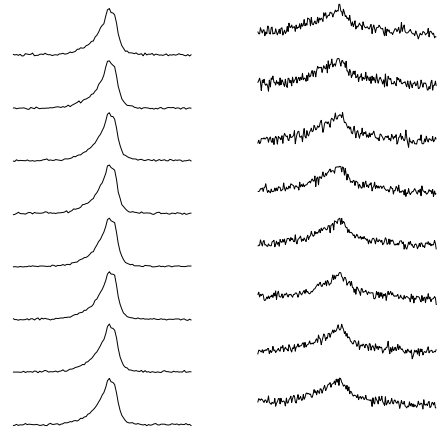} \\
    (a) \\
    \includegraphics[height=.3\textheight]{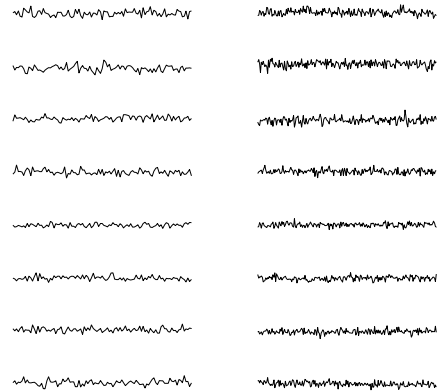}\\
    (b) \\

  \end{tabular}

  \caption{J1906+0746 pulse and interpulse for 8 orbital phase bins for two days of data are shown in the two leftmost sets of plots - a third day will be incorporated once it has been recalibrated. Note that the interpulse has been magnified by a factor of 8 relative to the main pulse. The two columns of plots on the right show the difference between the profile in each orbital bin and an average profile summed over the two days of data. The difference profiles show no evidence of profile changes on an orbital timescale within the noise. We will use these limits along with the long-term profile evolution to constrain the emission beam}
\end{figure}

\section{Timing}
We present timing using L-band data from telescopes at Green Bank, Arecibo and Nancay.  For each observation, we use the Gaussian template nearest in time.  A total of 1676 TOAs were fitted using the pulsar timing package TEMPO2 \cite{tempo2}.  Even with the multiple templates, there was still a large amount of timing noise.  We removed this using the 'fitwaves' feature of TEMPO2 \cite{hobbs}, which fit out 13 harmonically-related sinusoids as illustrated in Figure 4a, resulting in the residuals presented in Figure 4b.  Table 1 shows the preliminary orbital and mass parameters resulting from this fit.  More work is needed on the timing of this system since, for example, we still need to correct for frequency-dependent profile differences that are apparent in some of the Arecibo data.  From our timing analysis, we obtain an improved companion mass of 1.365$\pm$0.018 $M_\odot$, allowing us to more confidently assume that the companion is a neutron star.

\begin{figure}[h]
  \begin{tabular}{c}
    
    \includegraphics[height=.3\textheight]{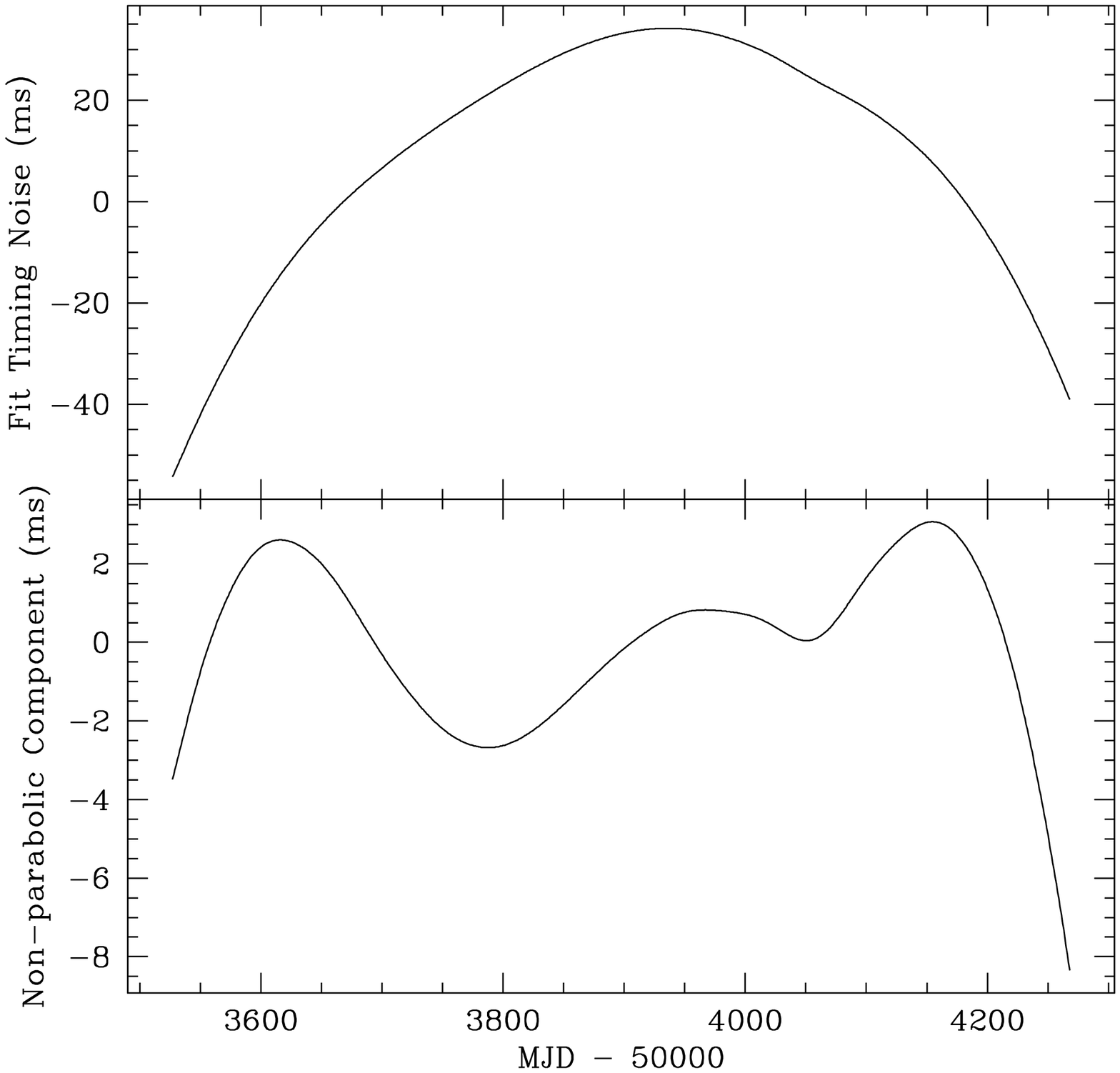} \\ 
    (a) \\
    \includegraphics[angle=90,angle=90,angle=90, height=.25\textheight]{resid} \\
    (b) \\

  \end{tabular}

  \caption{ a) The timing noise required 13 harmonically-related 'fitwaves' sinusoids to be fit out using TEMPO2, with a fundamental period of 2.322 years.  The resulting variations attributed to timing noise are plotted in the top panel.  In the bottom panel, we have subtracted from these the best-fit parabola to demonstrate the large short-term variations.  Based on the amplitude of the parabola, we estimate that the use of fitwaves has affected the fit period derivative at the level of about $2.5\%$. b) Timing residuals using TEMPO2 and a relativistic timing model (Damour $\&$ Deruelle 1986) \cite{damour}.  Residuals are black: Nancay; dark grey: Arecibo (WAPP); and light grey: GBT (GASP).}
\end{figure}

\begin{figure}[h]
  \includegraphics[height=.3\textheight]{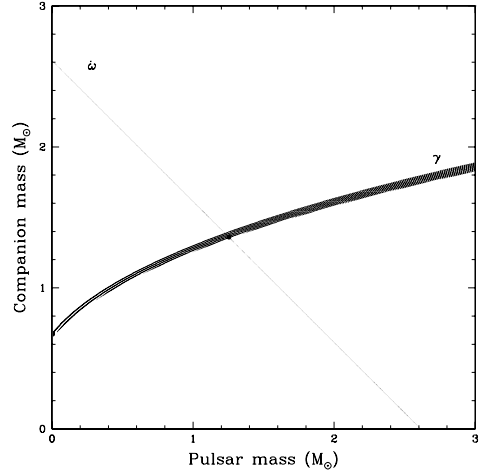}

  \caption{This mass--mass diagram displays constraints only from the advance of periastron $\dot\omega$ and the time dilation/gravitational redshift parameter $\gamma$. Adding the orbital period derivative to the fit changes the fit only marginally, and the fit does not yet appear to be robust. The black dot represents the masses derived assuming general relativity.}
\end{figure}

\begin{table}
\begin{tabular}{p{50pt}rlp{50pt}}
\hline
& $\dot\omega$ & 7.58331(95) &\\
& $\gamma$ & 0.0004930(84) & \\
& $M_1$ & 1.248(18) $M_\odot$ & \\
& $M_2$ & 1.365(18) $M_\odot$ & \\
\hline
\end{tabular}
\caption{Preliminary Mass and Orbital Parameters for J1906+0746}
\end{table}

\begin{theacknowledgments}
This work was supported by the NSF through a cooperative agreement with Cornell University to operate the Arecibo Observatory.  NRAO is a facility of the NSF operated under cooperative agreement by Associated Universities, Inc. L. K. holds an NSERC Canada Graduate Scholarship. 
\end{theacknowledgments}

%% The code below assumes that BibTeX is used. Compliant BibTex styles
%% are aipproc (for use with natbib) and aipprocl (if natbib is missing
%% at the site).
%%
%% Please run "bibtex \jobname" to obtain the bibliography and 
%% then re-run LaTeX twice to fix the references!
%%
%% When referring to citations in the text, in quare brackets [] show
%% the number in order of appearance. References in the References
%% section are listed in the same numerical order.
%%%%%%%%%%%%%%%%%%%%%%%%%%%%%%%%%%%%%%%%%%%%%%%%%

\bibliographystyle{aipprocl} % if natbib is missing


%%\bibliography{sample}

%%%%%%%%%%%%%%%%%%%%%%%%%%%%%%%%%%%%%%%%%%%%%%%%%
%% If the bibliography is
%% produced without BibTeX, comment out the above lines, use
%% \begin{thebibliography}{widest-label} environment to hold 
%% the list of references and 
%% \bibitem{label} command to start a bibliographical entry having
%% the "label" for use in \cite commands.

%%%%%%%%%%%%%%%%%%%%%%%%%%%%%%%%%%%%%%%%%%%
%% The following lines show an example how to produce a bibliography
%% without the help of the BibTeX program. This could be used instead
%% of the above.
%%%%%%%%%%%%%%%%%%%%%%%%%%%%%%%%%%%%%%%%%%%
\hyphenation{Post-Script Sprin-ger}

\end{document}